\documentclass[twoside]{IEEEtran}

\usepackage{amsmath,amssymb}
\usepackage{cite}
\usepackage{graphicx}
\usepackage[font=footnotesize]{caption}
\usepackage[font=footnotesize]{subcaption}

\newtheorem{thm}{Theorem}

\newtheorem{lem}[thm]{Lemma}

\newtheorem{myexample}{Example}
\newenvironment{exa}{\myexample}{\hfill\IEEEQED \endmyexample}
\newenvironment{dfn}{\textit{Definition: }}{}
\newenvironment{rem}{\textit{Remark: }}{}

\DeclareMathOperator{\rank}{rank}

\newcommand{\FF}{\mathbb{F}}

\newcommand{\ba}{\mathbf{a}}

\newcommand{\be}{\mathbf{e}}
\newcommand{\bg}{\mathbf{g}}
\newcommand{\bk}{\mathbf{k}}
\newcommand{\br}{\mathbf{r}}
\newcommand{\bu}{\mathbf{u}}
\newcommand{\bv}{\mathbf{v}}
\newcommand{\bx}{\mathbf{x}}
\newcommand{\by}{\mathbf{y}}

\newcommand{\bA}{\mathbf{A}}

\newcommand{\bE}{\mathbf{E}}
\newcommand{\bG}{\mathbf{G}}
\newcommand{\bR}{\mathbf{R}}
\newcommand{\bS}{\mathbf{S}}
\newcommand{\bT}{\mathbf{T}}
\newcommand{\bU}{\mathbf{U}}
\newcommand{\bV}{\mathbf{V}}
\newcommand{\bX}{\mathbf{X}}
\newcommand{\bY}{\mathbf{Y}}

\newcommand{\calJ}{\mathcal{J}}
\newcommand{\calK}{\mathcal{K}}
\newcommand{\calG}{\mathcal{G}}
\newcommand{\calP}{\mathcal{P}}
\newcommand{\calS}{\mathcal{S}}
\newcommand{\calT}{\mathcal{T}}
\newcommand{\calU}{\mathcal{U}}
\newcommand{\calV}{\mathcal{V}}
\newcommand{\calX}{\mathcal{X}}
\newcommand{\calY}{\mathcal{Y}}

\begin{document}

\title{On the Capacity of Multiplicative Finite-Field Matrix Channels}

\author{
  Roberto W. N\'obrega, \IEEEmembership{Student Member, IEEE}, Danilo Silva, \IEEEmembership{Member, IEEE}, and \\ Bartolomeu F. Uch\^oa-Filho, \IEEEmembership{Senior Member, IEEE}

  \thanks{This work was supported in part by CNPq--Brazil. The material in this paper was presented in part at the 2011 IEEE International Symposium on Information Theory~\cite{Nobrega.11.ISIT}. Some of the earlier ideas on which this work is based appeared in an unpublished draft~\cite{Uchoa.10.arXiv}.}

  \thanks{The authors are with the Department of Electrical Engineering of the Federal University of Santa Catarina, Florian\'opolis 88040--970, Brazil. (email: rwnobrega@eel.ufsc.br; danilo@eel.ufsc.br; uchoa@eel.ufsc.br).}

  \thanks{Copyright (c) 2013 IEEE. Personal use of this material is permitted.  However, permission to use this material for any other purposes must be obtained from the IEEE by sending a request to pubs-permissions@ieee.org.}
}

\maketitle

\begin{abstract}
  This paper deals with the multiplicative finite-field matrix channel, a discrete memoryless channel whose input and output are matrices (over a finite field) related by a multiplicative transfer matrix. The model considered here assumes that all transfer matrices with the same rank are equiprobable, so that the channel is completely characterized by the rank distribution of the transfer matrix. This model is seen to be more flexible than previously proposed ones in describing random linear network coding systems subject to link erasures, while still being sufficiently simple to allow tractability. The model is also conservative in the sense that its capacity provides a lower bound on the capacity of any channel with the same rank distribution. A main contribution is to express the channel capacity as the solution of a convex optimization problem which can be easily solved by numerical computation. For the special case of constant-rank input, a closed-form expression for the capacity is obtained. The behavior of the channel for asymptotically large field size or packet length is studied, and it is shown that constant-rank input suffices in this case. Finally, it is proved that the well-known approach of treating inputs and outputs as subspaces is information-lossless even in this more general model.
\end{abstract}

\begin{IEEEkeywords}
  Channel capacity, finite-field matrix channel, multiplicative matrix channel, noncoherent network coding, random linear network coding, subspace coding.
\end{IEEEkeywords}

\section{Introduction}

Finite-field matrix channels are communication channels where both the input and the output are matrices over some finite field $\FF_q$. The interest in such channels has been rising since the seminal work of Koetter and Kschischang~\cite{Koetter.Kschischang.08}, which connects finite-field matrix channels to the problem of error control in noncoherent network coding. In contrast with the combinatorial framework of~\cite{Koetter.Kschischang.08}, the present paper follows~\cite{Montanari.Urbanke.13,Silva.10,Jafari.11} and adopts a probabilistic approach.

The object of study of this work is the \emph{multiplicative finite-field matrix channel} (\emph{MMC}), modeled by the law\footnote{Throughout this paper, random entities are represented using boldface letters, while italic letters are used for their samples.}
\begin{equation} \label{eq:channel-law}
  \bY = \bG \bX,
\end{equation}
where $\bX \in \FF_q^{n \times \ell}$ is the channel \emph{input matrix}, $\bY \in \FF_q^{m \times \ell}$ is the channel \emph{output matrix}, and $\bG \in \FF_q^{m \times n}$ is the channel \emph{transfer matrix}, with $\bX$ and $\bG$ statistically independent. For simplicity, we assume $\max \{ n, m \} \leq \ell$.  This model turns out to be well-suited for random linear network coding systems~\cite{Ho.06} in the absence of malicious nodes, but possibly subject to link erasures. In this context, $\bX$ is the matrix whose rows are the $n$ packets transmitted by the source node, $\bY$ is the matrix whose rows are the $m$ packets received by the sink node, and $\ell$ is the number of $q$-ary symbols in each packet. Also, $\bG$~is the network transfer matrix, whose probability distribution is dictated by the network topology, the random choices of coding coefficients, and the link erasure probabilities.

Multiplicative finite-field matrix channels have been previously considered by Silva~{\it et al.}~\cite{Silva.10} and Jafari~{\it et al.}~\cite{Jafari.11}. Specifically, in~\cite{Silva.10}, $\bG$ is chosen uniformly at random among all full-rank matrices, while in~\cite{Jafari.11}, $\bG$ has i.i.d.\ entries selected uniformly at random (or, equivalently, $\bG$ is uniform over all matrices). Although these transfer matrix distributions could in principle be used to model random linear network coding systems, they cannot properly reflect different network topologies or accurately describe systems in which link erasures play an important role. This is because in these models the transfer matrix distribution is completely specified by the field size~$q$ and the dimensions~$n$ and~$m$. On the other hand, a full description of a completely general transfer matrix distribution requires, in addition, the specification of $q^{nm}$ parameters (namely, $\Pr [\bG = G]$, for $G \in \FF_q^{m \times n}$), therefore being impractical even for modest values of $q$, $n$ and~$m$.

In view of this tension between tractability and generality, the present paper suggests a new model which generalizes both the models of~\cite{Silva.10} and~\cite{Jafari.11}, but still keeps to a realistic level the amount of information needed to describe the channel.  Specifically, we allow the probability distribution of the \emph{rank} of $\bG$ to be arbitrary; nevertheless we consider that all matrices with the same rank are equiprobable.  We say such a transfer matrix is \emph{uniform given rank} (abbreviated as \emph{u.g.r.}). Under this assumption, the probability distribution of the rank of the transfer matrix completely determines the distribution of the transfer matrix itself and, therefore, also completely determines the channel.  Thus, the model only requires $\min \{n, m\} + 1$ parameters to describe the channel (namely, $\Pr [\rank \bG = r]$, for $0 \leq r \leq \min \{ n, m \}$).  While it is a challenging problem to obtain the rank distribution analytically for a general network topology (even in the simplest case of erasure-free links), in practice, a reasonable estimate may be obtained more simply by Monte Carlo simulation for a given network model. In fact, the (empirical) rank distribution is a natural figure of merit for most noncoherent network coding implementations (see, e.g.,~\cite{Chou.Wu.Jain.03.Allerton}). Thus, it is not entirely unrealistic to assume that this information is indeed available.

In order to convince the reader of the usefulness of the proposed model in practical scenarios, we provide an example (see Section~\ref{sec:examples}) on how the u.g.r.\ transfer matrix is able to better capture some properties of noncoherent network coding systems when compared to existing models. Specifically, we will see that for certain network topologies, the capacities in~\cite{Silva.10,Jafari.11} deviate more and more from the true capacity as the (graph) distance between the source and sink nodes increases or the link erasure probability grows. Furthermore, as we shall prove, any MMC can be reduced to our model (although with a potential decrease in the channel capacity) by means of a simple preprocessing at the transmitter and receiver.  Since this preprocessing does not alter the rank distribution of the transfer matrix, this implies that among all transfer matrices sharing the same rank distribution, the u.g.r.\ is the one with lowest channel capacity.  In this sense, the u.g.r.\ model seems to arise naturally in the study of multiplicative finite-field matrix channels.

In this paper, we concentrate on the problem of finding the capacity and mutual information of the MMC with u.g.r.\ transfer matrix. We show that the capacity is achieved when the input matrix (similarly to the transfer matrix) is u.g.r., and an expression for the mutual information is derived for this kind of input. As a consequence, we are able to greatly reduce the complexity of the convex optimization problem involved in obtaining the channel capacity and the associated optimal input, when compared to the most general MMC model---a reduction from $q^{n\ell}$ to $n + 1$ variables, as we shall see. We then turn over to the special situation of constant-rank input.  In this case, we are able to obtain a closed-form expression for the constant-rank capacity. Later on, we consider the problem in which $q$ or $\ell$ are allowed to grow arbitrarily, and show that the true channel capacity is achieved by constant-rank input. As a final contribution, we verify that communication via subspaces is still optimal when the transfer matrix is u.g.r.  This generalizes similar conclusions previously obtained in~\cite{Silva.10} and~\cite{Jafari.11}.

A related line of work by Yang {\it et al.}~\cite{Yang.10.arXiv,Yang.11.arXiv,Yang.10.ITW,Yang.10.ISIT}, done concurrently to and independently of our work, considers a completely general transfer matrix distribution (with the transfer matrix still independent of the input).  They were able to identify a class of inputs (which they call ``$\alpha$-type'') that is sufficient to achieve the channel capacity. As a result, the number of optimization variables required to compute the channel capacity is reduced---although to a number that is still exponential in the matrix size. They also derive upper and lower bounds on the capacity which depends only on the rank distribution of the transfer matrix. It is worth mentioning that some of our results can be obtained by specializing the results in~\cite{Yang.10.arXiv} to a u.g.r.\ transfer matrix. (Appropriate comparisons are made along the text whenever applicable.) Nevertheless, we believe that the approach we follow here is simpler and more insightful for this particular case.

Finally, it is worth noticing that some of the results obtained in this paper have been subsequently employed in~\cite{Jafari.Yang.12}, where an \emph{arbitrarily varying channel} approach to the MMC is considered. More precisely,~\cite{Jafari.Yang.12} assumes that the rank of the transfer matrix is randomly chosen according to a known probability distribution, but, apart from that, the transfer matrix can be changed arbitrarily from time-slot to time-slot. It is shown that the capacity of this channel is the same as the capacity of the MMC with u.g.r.\ transfer matrix considered here.

The remainder of this paper is organized as follows. Section~\ref{sec:background} presents some notation, basic facts, and a brief review on discrete memoryless channels. Section~\ref{sec:channel-model} defines the channel model under consideration. Section~\ref{sec:examples} considers a motivating example. Section~\ref{sec:results} contains the main results of this work, whose proofs are located in Section~\ref{sec:proofs}. Section~\ref{sec:conclusion} concludes the paper.

\section{Notation and Background}\label{sec:background}

Let $\FF_q$ be a finite field. We denote by $\FF_q^{m \times n}$ the set of all $m \times n$ matrices with entries in $\FF_q$, and by $\calT_r(\FF_q^{m \times n})$ those matrices in $\FF_q^{m \times n}$ with rank~$r$. For notational convenience, we sometimes set $\calT_r = \calT_r(\FF_q^{m \times n})$ when the matrix dimension~$m \times n$ and the field size~$q$ are implied by the context. Also, $\calT(\FF_q^{m \times n}) \triangleq \calT_{\min \{ n, m \}}(\FF_q^{m \times n})$ is the set of all $m \times n$ full-rank matrices. It is well-known (see, e.g.,~\cite{Fisher.66}) that
\[
  |\calT(\FF_q^{m \times r})| = \prod_{i=0}^{r-1}(q^m - q^i),
\]
for $r \leq m$, and
\begin{equation} \label{eq:chi-rank}
  |\calT_r(\FF_q^{m \times n})| = |\calT(\FF_q^{m\times r})| {n \brack r}_q,
\end{equation}
where
\begin{equation} \label{eq:gaussian-binomial}
  {n \brack r}_q \triangleq
  \begin{cases}
    \displaystyle \prod_{i=0}^{r-1}\frac{q^n-q^i}{q^r-q^i}, & \text{if } 0 \leq r \leq n, \\
    0, & \text{else,}
  \end{cases}
\end{equation}
denotes the Gaussian binomial coefficient. It is also known that the Gaussian binomial coefficient satisfies~\cite[Lemma~4]{Koetter.Kschischang.08}
\begin{equation} \label{eq:gaussian-binomial-bounds}
  q^{r(n - r)} \leq {n \brack r}_q \leq \gamma_q \, q^{r(n - r)},
\end{equation}
where
\[
  \gamma_q = \prod_{i=1}^\infty \frac 1 {1 - q^{-i}}.
\]

In this paper, we let $\langle A \rangle$ denote the row space of a matrix~$A$, and $1[P]$ the indicator function of $P$, that is,
\[
  1[P] =
  \begin{cases}
    1, & \text{if } P \text{ is true,} \\
    0, & \text{otherwise.}
  \end{cases}
\]

A \emph{discrete memoryless channel} (\emph{DMC})~\cite{Cover.06} with input $\bx$ and output $\by$ is defined by a triplet $(\calX,p_{\by|\bx},\calY)$, where $\calX$ and $\calY$ are the channel input and output alphabets, respectively, and $p_{\by|\bx}$, called the \emph{channel transition probability}, gives the conditional probability that $\by = y \in \calY$ is received given that $\bx = x \in \calX$ is sent. The channel is \emph{memoryless} in the sense that what happens to the transmitted symbol at one time is independent of what happens to the transmitted symbol at any other time. The \emph{capacity} of the DMC is then given by
\[
  C = \max_{p_\bx} I(\bx ; \by),
\]
where $I(\bx;\by)$ is the mutual information between $\bx$ and $\by$, and the maximization is over all possible input distributions~$p_\bx$.

An interesting question is whether input or output letters of a DMC can be grouped together without reducing the channel mutual information. The following result (see, e.g., \cite[\S 5.9--5.10]{Abramson.63}) derives the conditions under which such groupings are information-lossless.

\medskip

\begin{lem} \label{lem:grouping}
  Let $(\calX, p_{\by|\bx}, \calY)$ be a DMC with input $\bx$ and output $\by$.  In addition, let $f : \calX \to \calU$ and  $g : \calY \to \calV$ be surjective functions, and define $\bu = f(\bx)$ and $\bv = f(\by)$. The following holds:
  \begin{enumerate}
    \item $I(\bx ; \by) = I(\bu ; \by)$ for all $p_\bx$ if and only if, for every pair $x, x' \in \calX$ satisfying $f(x) = f(x')$, we have $p_{\by|\bx}(y|x) = p_{\by|\bx}(y|x')$ for all $y \in \calY$.
    \item $I(\bx ; \by) = I(\bx ; \bv)$ for all $p_\bx$ if and only if, for every pair $y, y' \in \calY$ satisfying $g(y) = g(y')$, there exists some real number $\alpha$ such that $p_{\by|\bx}(y'|x) = \alpha \, p_{\by|\bx}(y|x)$ for all $x \in \calX$.
  \end{enumerate}
\end{lem}

\section{Channel Model} \label{sec:channel-model}

The MMC described by the channel law~\eqref{eq:channel-law} can naturally be viewed as a DMC defined by
\[
 (\calX = \FF_q^{n \times \ell}, \ p_{\bY|\bX}, \ \calY = \FF_q^{m \times \ell}),
\]
where the channel transition probability is given by
\begin{align*}
  p_{\bY|\bX}(Y|X)
    & = \sum_G p_{\bG|\bX}(G|X) \, p_{\bY|\bX,\bG}(Y|X,G) \\
    & = \sum_G p_\bG(G) \, 1[Y = GX]
\end{align*}
(and thus completely characterized by~$p_\bG$). This work deals with a special class of this channel, in which the transfer matrix $\bG$ is ``uniform given rank,'' a concept defined next.

\medskip

\begin{dfn}
  A random matrix $\bA \in \FF_q^{m \times n}$ distributed~according to $p_\bA$ is said to be \emph{uniform given rank} (\emph{u.g.r.}, for~short) if, for every $A,A' \in \FF_q^{m \times n}$, we have $p_\bA(A) = p_\bA(A')$ whenever $\rank A = \rank A'$.
\end{dfn}

\medskip

Let $\bA$ be a random matrix over $\FF_q^{m \times n}$ with probability distribution~$p_\bA$. Also, let $\bk = \rank \bA$; this is a random variable taking values on $\{ 0, \ldots, \min \{ n, m \} \}$ according to a probability distribution~$p_\bk$ given by
\[
  p_\bk(k) = \sum_{A \in \calT_k} p_\bA(A).
\]
Then, it is clear that $\bA$ is u.g.r.\ if and only if
\[
  p_\bA(A) = \frac{p_\bk(k)} {|\calT_k(\FF_q^{m \times n})|},
\]
where $k = \rank A$. In this way, the rank probability distribution $p_\bk$ completely determines $p_\bA$ for $\bA$ u.g.r.  In addition, it is not hard to show that the entropy of~$\bA$ satisfies
\begin{equation} \label{eq:entropy-bound}
  H(\bA) \leq \sum_k p_\bk(k) \log_q \frac {|\calT_k(\FF_q^{m \times n})|} {p_\bk(k)},
\end{equation}
with equality when $\bA$ is u.g.r.  This is because among all matrices with a given rank probability distribution, the u.g.r.\ is the one with largest entropy.

As said before, both the models of Silva~{\it et al.} \cite{Silva.10} and Jafari~{\it et al.} \cite{Jafari.11} are special cases of the u.g.r. model considered here.  Indeed, let $\br = \rank\bG$, distributed according to  $p_\br(r) = \sum_{G \in \calT_r} p_\bG(G)$, be the random variable representing the rank of the transfer matrix.  Then, for the channel model in~\cite{Silva.10}, where $\bG$ is uniformly distributed over~$\calT(\FF_q^{n \times n})$, we have
\begin{equation} \label{eq:rank-silva}
  p_\br(r) =
  \begin{cases}
    1, & \text{if } r = n, \\
    0, & \text{else,}
  \end{cases}
\end{equation}
while for the channel model in~\cite{Jafari.11}, where $\bG$ is uniformly distributed over $\FF_q^{m \times n}$, we have
\begin{equation} \label{eq:rank-jafari}
  p_\br(r) = \frac{|\calT_r(\FF_q^{m \times n})|}{q^{nm}}.
\end{equation}

We remark that every MMC can be artificially transformed into an MMC with u.g.r.\ transfer matrix (having the same rank distribution as the original channel) by means of ``randomization'' at both the transmitter and receiver. Theorem~\ref{thm:worst-case} below makes this precise.  We prove this theorem as an application of a generalized version of the crypto lemma~\cite{Forney.03.Allerton}, which may be useful in other applications. The proofs are given in Appendix~\ref{ape:crypto}.

\medskip

\begin{thm}\label{thm:worst-case}
  Let $\bG \in \FF_q^{m \times n}$ be a random matrix with arbitrary probability distribution, and define $\bG' = \bT_1 \bG \bT_2$, where $\bT_1 \in \calT(\FF_q^{m \times m})$ and $\bT_2 \in \calT(\FF_q^{n \times n})$ are uniformly distributed full-rank square matrices, independent of~$\bG$ and of each other. Then, $\bG'$ is u.g.r.\ and has the same rank distribution as~$\bG$.
\end{thm}

\medskip

\begin{figure}
  \centering
  \includegraphics[width=7.5cm]{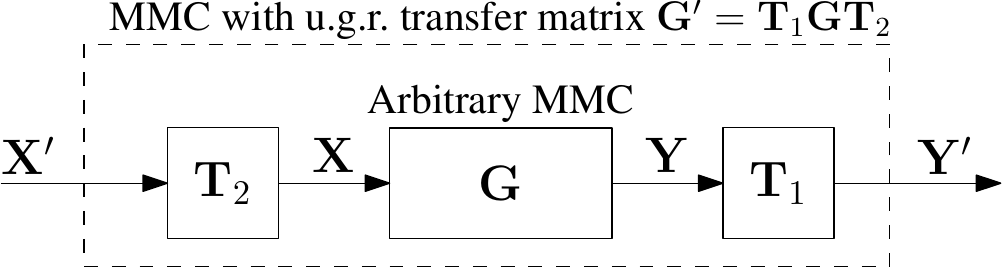}
  \caption{Turning an arbitrary MMC into an MMC with u.g.r.\ transfer matrix. The rank distribution of the new channel is the same as the original channel.}
  \label{fig:worst-case}
\end{figure}

Effectively (see Fig.~\ref{fig:worst-case}), instead of transmitting the original source packets (say $\bX'$), the transmitter sends $\bX = \bT_2 \bX'$; and instead of the actual channel output (say $\bY$), the receiver considers $\bY' = \bT_1 \bY$ for decoding. (Here, $\bT_1$ and $\bT_2$ are defined as in Theorem~\ref{thm:worst-case}.) Consequently, if the transfer matrix of the original channel is~$\bG$, we have $\bY' = \bT_1 \bY = \bT_1 \bG \bX = \bT_1 \bG \bT_2 \bX' = \bG' \bX'$, where~$\bG'$, according to Theorem~\ref{thm:worst-case}, is u.g.r. and has the same rank distribution as $\bG$. Naturally, from the data-processing inequality~\cite{Cover.06}, we have $I(\bX' ; \bY') \leq I(\bX ; \bY)$, so that this transformation comes at the expense of a potential reduction of the channel capacity.

Thus, we conclude that, among all transfer matrices sharing the same rank distribution, the u.g.r.\ is the one with lowest channel capacity, and that any capacity result obtained for the MMC with u.g.r.\ transfer matrix can be used as a lower bound for MMCs with non-u.g.r.\ transfer matrices.

A few more comments are in order. First, note that randomization at the transmitter (but not at the receiver) is already a usual practice in random linear network coding systems~\cite{Koetter.Kschischang.08}. Second, since both the multiplication of matrices and the generation of a random invertible matrix can be accomplished in polynomial time, the randomization is also a polynomial-time procedure.  Third, because $\bT_1$ and $\bT_2$ are independent of $\bG$ and of each other, no channel knowledge is assumed, and no common randomness shared by the transmitter and receiver is required.  Finally, for a numerical quantification of the rate loss incurred by randomization, refer to Example~\ref{exa:w-layered-2x2} in Section~\ref{sec:examples}.

\section{Motivating Example}\label{sec:examples}

\begin{figure}
  \centering
  \includegraphics[width=8cm]{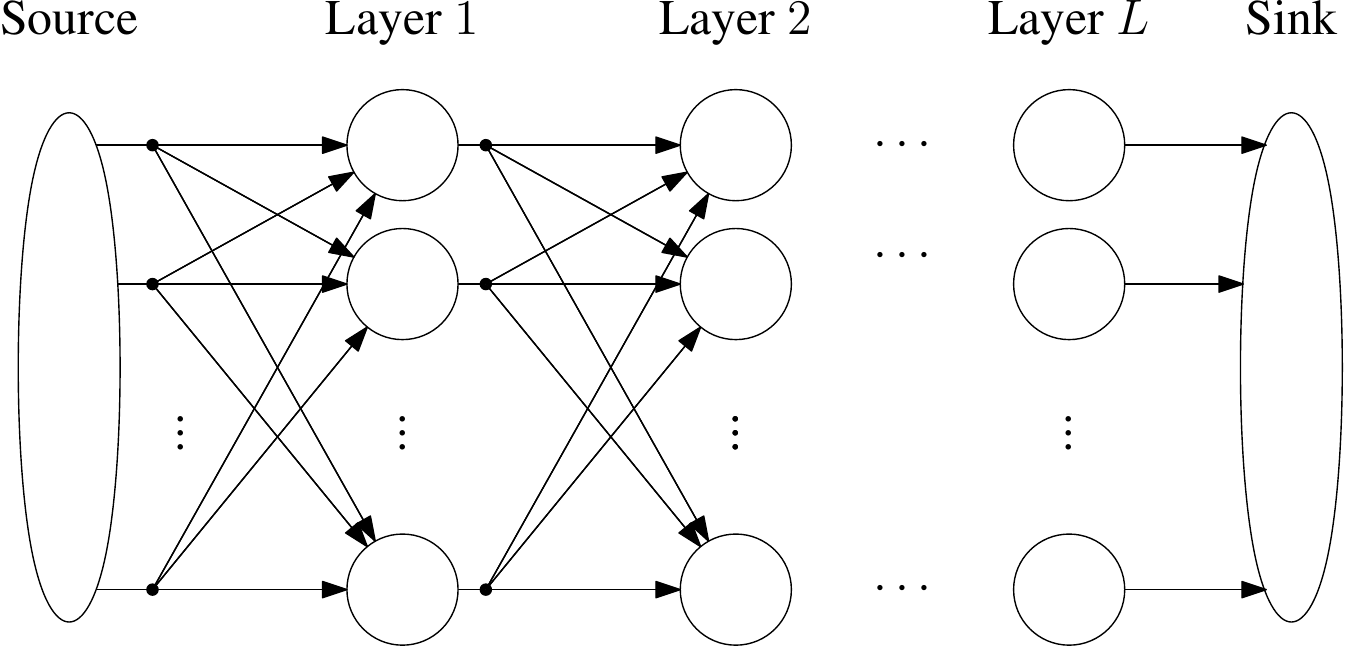}
  \caption{Wireless layered relay network. There are $L$ layers, and each layer has $N$ relay nodes.}
  \label{fig:w-layered-topology}
\end{figure}

In this section, we present an example showing how the u.g.r.\ model is able to better model a noncoherent network coding system. Consider the wireless relay network depicted in Fig.~\ref{fig:w-layered-topology}, with $L$ layers (columns) and $N$ relay nodes per layer. Assume that the system operates with packets of length~$\ell$, and that between each two consecutive layers (also between the source node and layer~$1$, and layer~$L$ and the sink node) there are $N$ orthogonal broadcast channels, which are subject to independent erasures occurring in the end of the channel with probability~$\epsilon$. Whenever a packet is erased, it is considered to be received as the all-zero vector. In addition, assume that there is no communication between nonadjacent layers, as well as between nodes in the same layer.

The system operates as follows. First, the source node transmits packets to the first layer by using all the $N$ orthogonal broadcast channels. It repeats this process $M$ times, so that a total of $MN$ packets is received by each node in the first layer. (It is assumed that the source does not perform any randomization.) After that, each node in the first layer computes $M$ random linear combinations (with i.i.d.\ uniform coefficients in $\FF_q$) of all its received packets, and broadcasts these linear combinations to the second layer, again in $M$ time slots, by using one of the $N$ orthogonal channels assigned to it. In this way, a total of $MN$ packets is received by each node in the second layer, $M$ from each node of the first layer. The system operates similarly up to layer~$L$. Finally, the sink node receives $MN$ packets, $M$ from each node in layer~$L$.

We now show that this system can be modeled as an MMC with $n = m = MN$. Let $\bX \in \FF_q^{M\!N \times \ell}$ (resp., $\bY \in \FF_q^{M\!N \times \ell}$) denote the matrix whose rows are the packets transmitted (resp., received) by the source (resp., sink) node. Let $\bR_{i,j} \in \FF_q^{M\!N \times \ell}$ (resp., $\bS_{i,j} \in \FF_q^{M \times \ell}$) denote the matrix whose rows are the packets received (resp., transmitted) by the $j$-th relay node of the $i$-th layer, for $1 \leq i \leq L$ and $1 \leq j \leq N$. From the network operation just described, we know that
\[
  \bS_{i,j} = \bA_{i,j} \bR_{i,j},
\]
for $1 \leq i \leq L$ and $1 \leq j \leq N$, where $\bA_{i,j} \in \FF_q^{M \times M\!N}$ are matrices whose entries are i.i.d.\ selected uniformly at random. We also know that
\[
  \bR_{1,j} = \bE_{1,j} \bX,
\]
\[
  \bR_{i,j} = \bE_{i,j} \left[
  \begin{array}{c}
    \bS_{i-1,1} \\
    \vdots \\
    \bS_{i-1,N} \\
  \end{array}
  \right],
  \quad \text{and} \quad
  \bY = \bE' \left[
  \begin{array}{c}
    \bS_{L,1} \\
    \vdots \\
    \bS_{L,N} \\
  \end{array}
  \right],
\]
for $2 \leq i \leq L$ and $1 \leq j \leq N$, where $\bE_{i,j}, \bE' \in \FF_q^{M\!N \times M\!N}$ are diagonal matrices (modeling the erasures) whose diagonal entries are i.i.d.\ with $p(0) = \epsilon$ and $p(1) = 1 - \epsilon$. From this, we can deduce that $\bY = \bG \bX$, where
\begin{equation} \label{eq:example-transfer-matrix}
  \bG = \bE' \bA_L \bE_L \cdots \bA_2 \bE_2 \bA_1 \bE_1,
\end{equation}
in which $\bA_i \in \FF_q^{M\!N \times M\!N^2}$ (a block-diagonal matrix) and $\bE_i \in \FF_q^{M\!N^2 \times M\!N}$ are given by
\[
  \bA_i = \left[
  \begin{array}{ccc}
    \bA_{i,1} &  & \\
    & \ddots & \\
    & & \bA_{i,N}
  \end{array}
  \right],
  \quad
  \bE_i = \left[
  \begin{array}{c}
    \bE_{i,1} \\
    \vdots \\
    \bE_{i,N}
  \end{array}
  \right].
\]

Note that, in general, the transfer matrix given in~\eqref{eq:example-transfer-matrix} is \emph{not} u.g.r.  Therefore, as mentioned in Section~\ref{sec:channel-model}, the capacity results from Section~\ref{sec:results} will serve only as lower bounds on the channel capacity. We herein call the attention to the fact that the calculation of the real value of the channel capacity is a computationally heavy task, even for small values of parameters. For example, when $q=2$ and $n=m=\ell=8$, a priori, we need to solve an optimization problem over $q^{n \ell} = 2^{64}$ variables, which is clearly impractical. According to~\cite{Yang.10.arXiv}, we could simplify the problem to $\sum_{u=0}^n {n \brack u}_q > 2^{18}$ variables, but this number is still impractical.

\begin{figure*}
  \centering
  \begin{subfigure}[b]{0.45\textwidth}
    \centering
    \includegraphics[width=8cm]{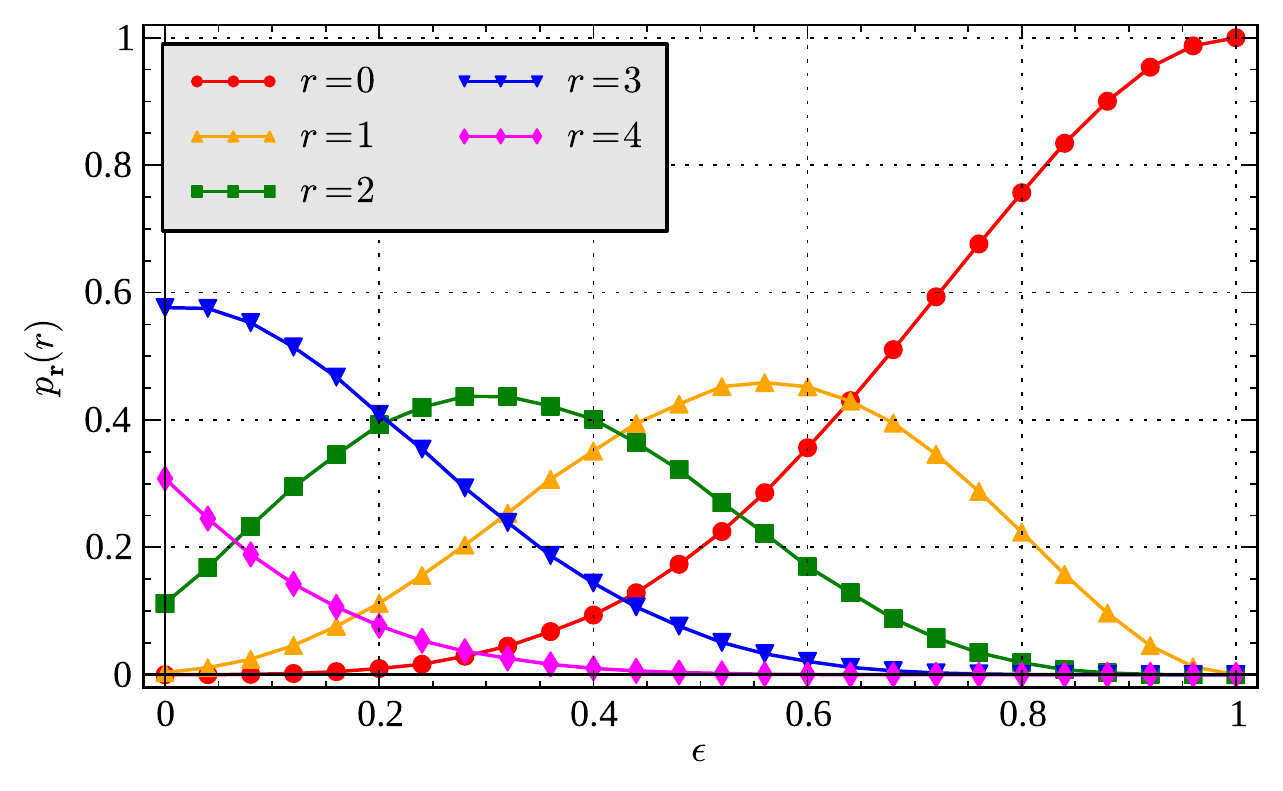}
    \caption{Rank distribution for $L = 1$, as a function of $\epsilon$.}
    \label{fig:w-layered-8x8-pr-epsilon}
  \end{subfigure}%
  \quad
  \begin{subfigure}[b]{0.45\textwidth}
    \centering
    \includegraphics[width=8cm]{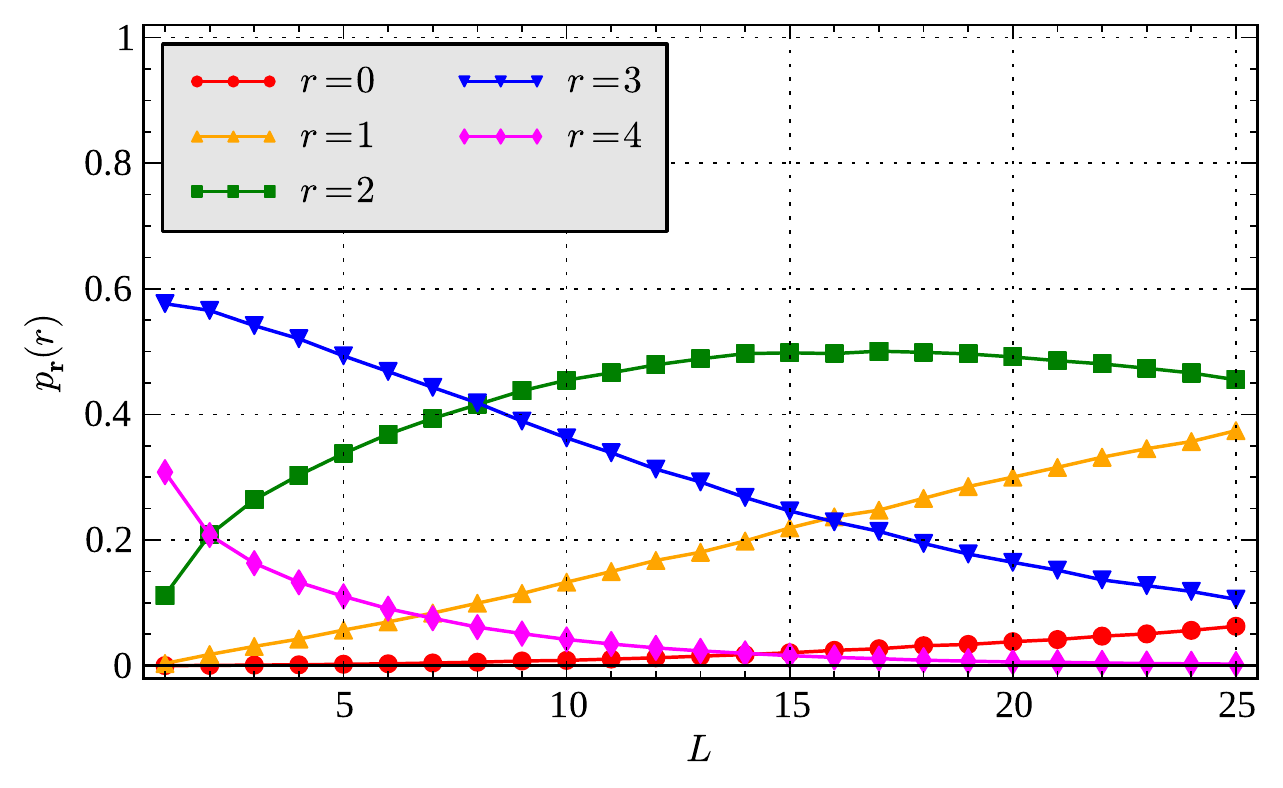}
    \caption{Rank distribution for $\epsilon = 0$, as a function of $L$.}
    \label{fig:w-layered-8x8-pr-depth}
  \end{subfigure}

  ~

  \begin{subfigure}[b]{0.45\textwidth}
    \centering
    \includegraphics[width=8cm]{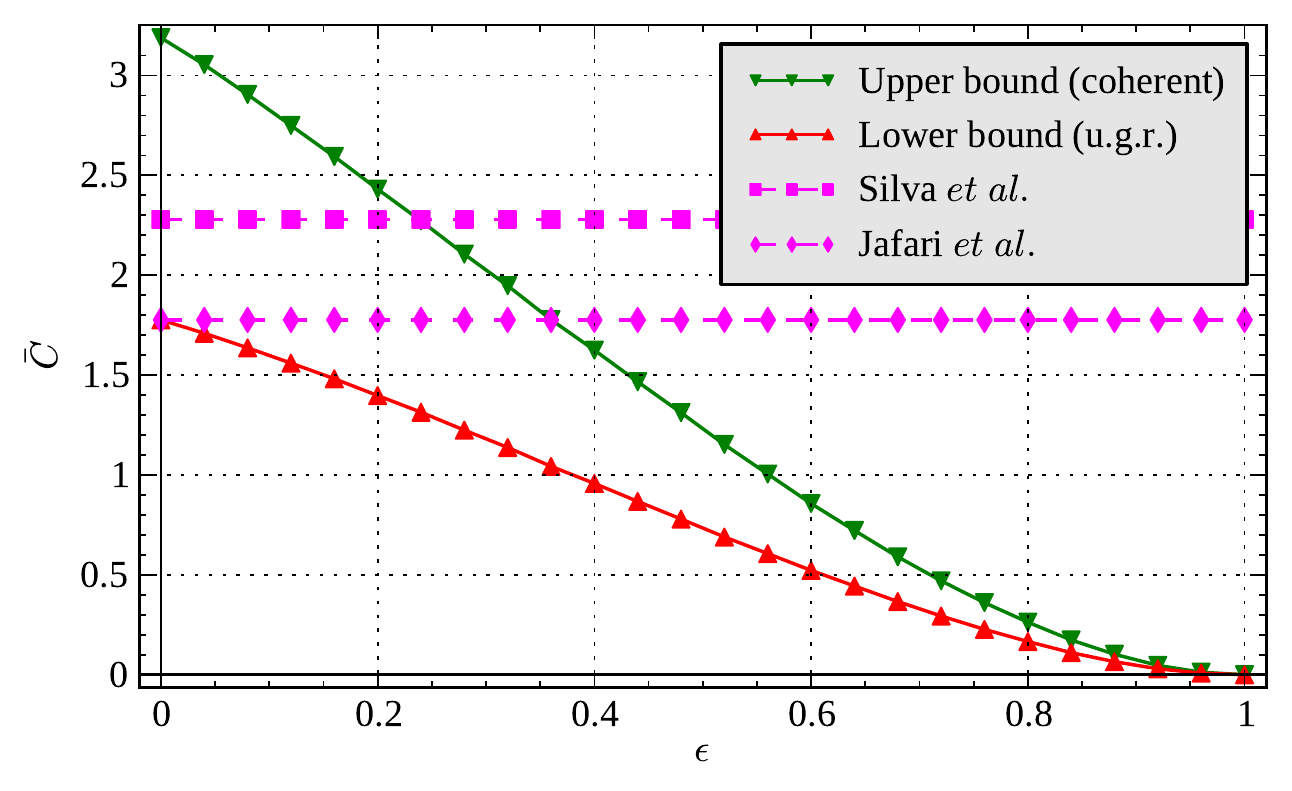}
    \caption{Capacity for $L = 1$ and $\ell = 8$, as a function of $\epsilon$.}
    \label{fig:w-layered-8x8-cap-epsilon}
  \end{subfigure}%
  \quad
  \begin{subfigure}[b]{0.45\textwidth}
    \centering
    \includegraphics[width=8cm]{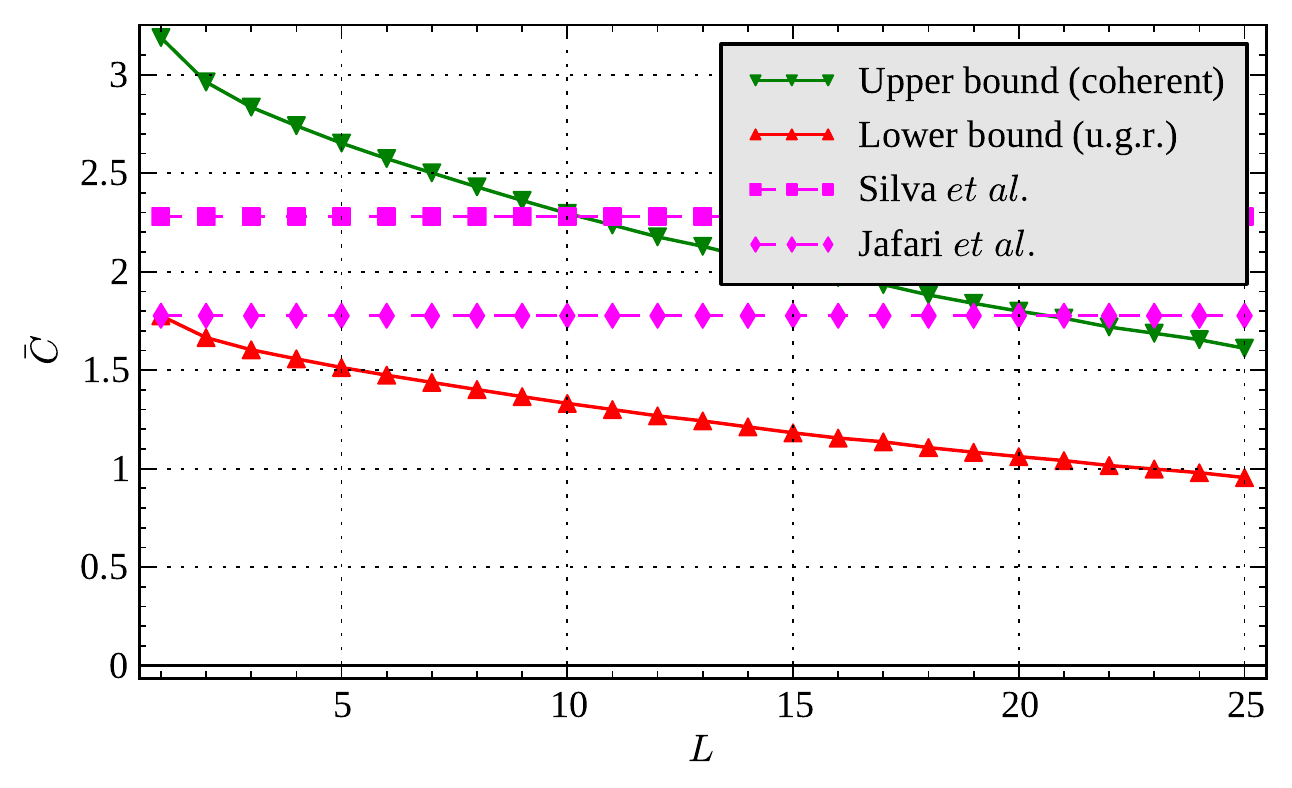}
    \caption{Capacity for $\epsilon = 0$ and $\ell = 8$, as a function of $L$.}
    \label{fig:w-layered-8x8-cap-depth}
  \end{subfigure}
  \caption{Rank distribution and channel capacity for the wireless layered relay network with $N = M = 2$ and $q = 2$.}
\end{figure*}

\medskip

\begin{exa}
  Figs.~\ref{fig:w-layered-8x8-pr-epsilon} and~\ref{fig:w-layered-8x8-pr-depth} show the rank distribution~$p_\br$ induced by the wireless layered relay network with $q = 2$ and $N = M = 2$ (thus, $n = m = MN = 4$), as a function of $\epsilon$, for $L = 1$, and as a function of $L$, for $\epsilon = 0$, respectively. Note that the value of $\ell$ is unimportant here.  Both rank distributions were obtained from~\eqref{eq:example-transfer-matrix} by the Monte~Carlo method with 100,000 realizations.

  Figs.~\ref{fig:w-layered-8x8-cap-epsilon} and~\ref{fig:w-layered-8x8-cap-depth} show the channel capacity of the corresponding MMC assuming u.g.r.\ transfer matrix, with the rank distributions of Figs.~\ref{fig:w-layered-8x8-pr-epsilon} and~\ref{fig:w-layered-8x8-pr-depth}, and considering a packet length $\ell = 8$. The results were obtained from Theorem~\ref{thm:capacity} of Section~\ref{sec:results}. The figures also show the capacity obtained for a system with the same parameters $q$, $n$, $m$, and $\ell$, but modeled with a full-rank uniform transfer matrix~\cite{Silva.10} or with a uniform transfer matrix~\cite{Jafari.11}, as well as the coherent upper bound of~\cite{Yang.10.arXiv} (i.e., the channel capacity assuming that both the transmitter and receiver know the transfer matrix).
\end{exa}

\medskip

Clearly, the models of~\cite{Silva.10} and~\cite{Jafari.11} are insensitive to the effects of link erasures and variations on the topology (here illustrated by the number of layers). The capacities for theses models are seen to deviate substantially from the true capacity. In contrast, from the trends of the lower and upper bounds curves, it can be inferred that the capacity for the u.g.r.\ model behaves much like the true capacity (note that the upper bound goes to zero as $\epsilon$ approaches one or $L$ increases; therefore, so does the true capacity). In fact, as the next example illustrates, the u.g.r.\ lower bound may actually be close to the true capacity.

\begin{figure*}
  \centering
  \begin{subfigure}[b]{0.45\textwidth}
    \centering
    \includegraphics[width=8cm]{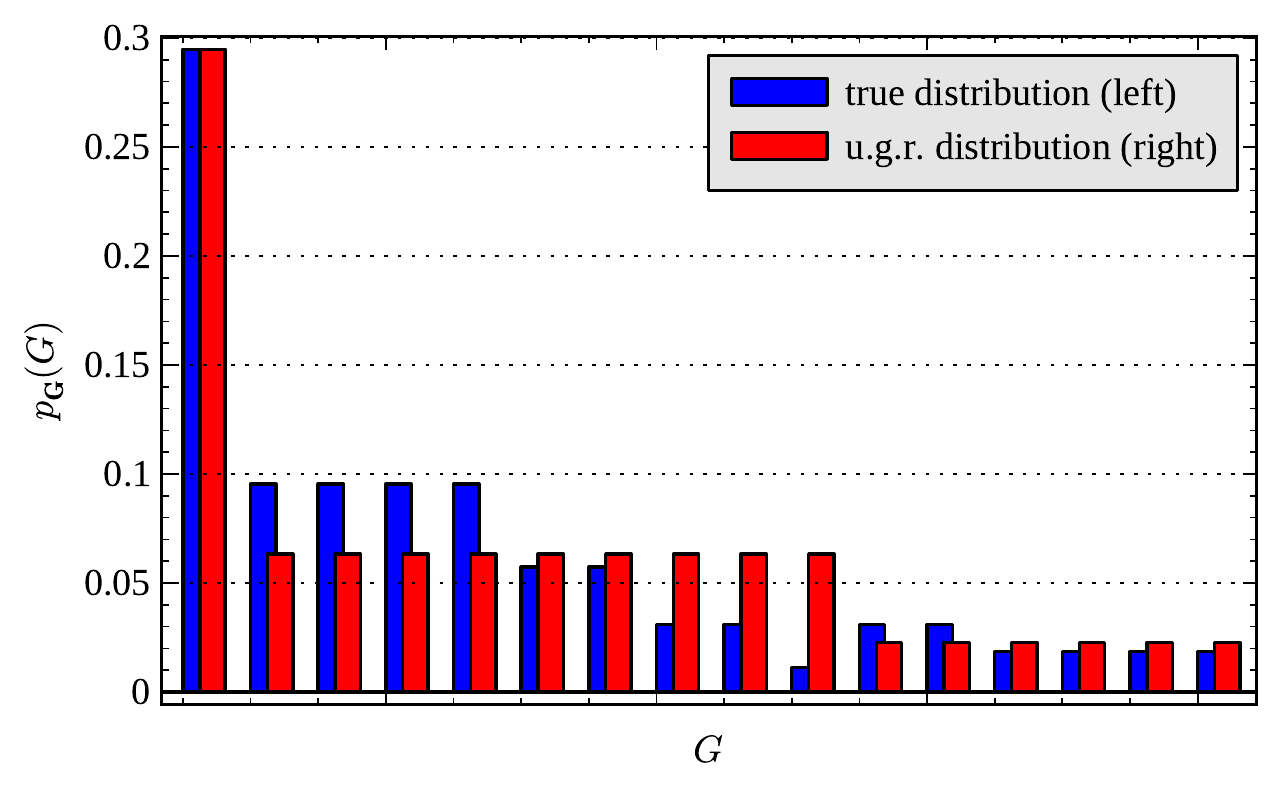}
    \caption{True and u.g.r.\ transfer matrix distributions, for $\epsilon = 1/4$.}
    \label{fig:w-layered-2x2-pG}
  \end{subfigure}%
  \quad
  \begin{subfigure}[b]{0.45\textwidth}
    \centering
    \includegraphics[width=8cm]{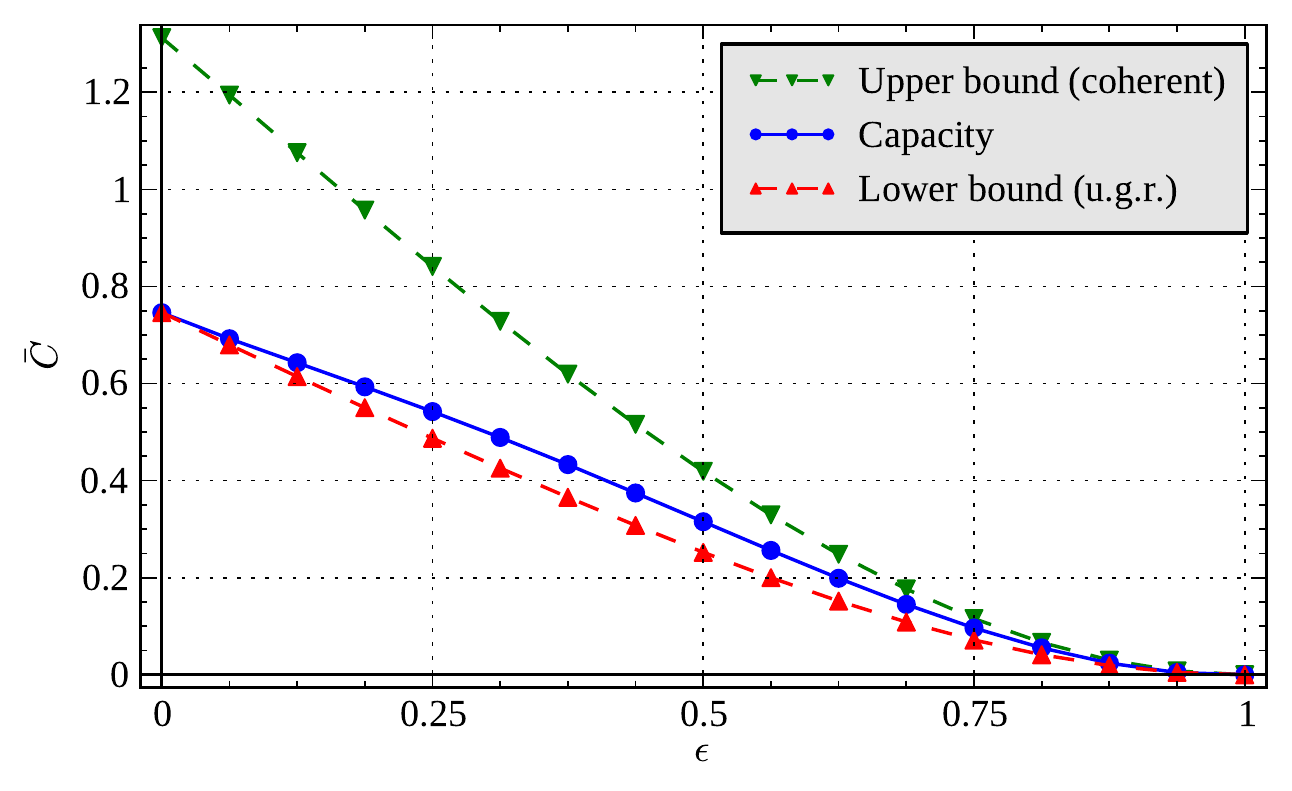}
    \caption{True capacity and bounds, as a function of $\epsilon$, for $\ell=3$.}
    \label{fig:w-layered-2x2-cap-epsilon}
  \end{subfigure}
  \caption{Transfer matrix distribution and channel capacity for the wireless layered relay network with $L=1$, $N=2$, $M=1$, and $q=2$. In Fig.~\ref{fig:w-layered-2x2-pG}, the horizontal axis consists of all the matrices in $\FF_2^{2 \times 2}$, ordered from left to right as follows: $[00;00]$, $[10;00]$, $[01;00]$, $[00;10]$, $[00;01]$, $[11;00]$, $[00;11]$, $[10;10]$, $[01;01]$, $[11;11]$, $[10;01]$, $[01;10]$, $[11;10]$, $[11;01]$, $[10;11]$, $[01;11]$.}
\end{figure*}

\medskip

\begin{exa} \label{exa:w-layered-2x2}
  This example aims to quantify the rate loss incurred by considering a matrix channel being u.g.r. when, in fact, it is not. For such, we consider the wireless layered relay network with field size $q=2$, a single layer ($L=1$), and two relay nodes ($N=2$). We also set $M=1$, so that $n=m=2$. In this case,~\eqref{eq:example-transfer-matrix} yields
  \[
    \bG = \bE' \bA_1 \bE_1 =
    \left[
    \begin{array}{cc}
      \be_5 \ba_1 \be_1 & \be_5 \ba_2 \be_2 \\
      \be_6 \ba_3 \be_3 & \be_6 \ba_4 \be_4 \\
    \end{array}
    \right],
  \]
  where $\be_1, \ldots, \be_6 \in \FF_2$ (related to the erasures) are i.i.d.\ with $\Pr[\be_i=0]=\epsilon$, and $\ba_1, \ldots, \ba_4 \in \FF_2$ (the network coding coefficients) are i.i.d.\ with $\Pr[\ba_i=0]=1/2$. The transfer matrix distribution $p_\bG(G)$ with $\epsilon = 1/4$ is shown in Fig.~\ref{fig:w-layered-2x2-pG}, which also shows the corresponding u.g.r.\ distribution.

  Fig.~\ref{fig:w-layered-2x2-cap-epsilon} shows the true channel capacity (obtained by solving the original maximization problem over $q^{n\ell} = 64$~variables), along with the u.g.r.\ lower bound (obtained by solving a maximization problem over $n+1 = 3$ variables, according to Theorem~\ref{thm:capacity}), and the coherent upper bound (given by $E[\br]$, according to Yang \textit{et al.}~\cite{Yang.10.arXiv}), as a function of $\epsilon$, for a packet length $\ell=3$. It is interesting to observe that the u.g.r. lower bound is tight for $\epsilon=0$, since in this case $\bG$~becomes uniformly distributed over $\FF_q^{m \times n}$, and thus u.g.r. Also, for all other values of $\epsilon$, the true capacity is very close to the u.g.r.\ lower bound, which constitutes an evidence that the u.g.r.\ model serves as a good approximation for noncoherent network coding systems.
\end{exa}

\section{Main Results}\label{sec:results}

This section present the main results of this work, whose proofs are left to Section~\ref{sec:proofs}. In what follows, we consider an MMC with input matrix~$\bX$, output matrix~$\bY$, and  u.g.r.\ transfer matrix~$\bG$. In addition to $\br\triangleq\rank\bG$, distributed according to $p_\br(r) = \sum_{G \in \calT_r} p_\bG(G)$, we also make use of the random variables $\bu\triangleq\rank\bX$ and $\bv\triangleq\rank\bY$, whose probability distributions are given by $p_\bu(u) = \sum_{X \in \calT_u} p_\bX(X)$ and $p_\bv(v) = \sum_{Y \in \calT_v} p_\bY(Y)$, respectively.

The \emph{rank transition probability}, that is, the probability of receiving a matrix with rank $\bv = v$ given the transmitted matrix has rank $\bu = u$, plays an important role in this work. Since $\bu \to \bX \to \bY \to \bv$ forms a Markov chain, the rank transition probability is given by
\begin{align*}
  p_{\bv|\bu}(v|u)
    & = \sum_{X,Y} p_{\bv|\bY}(v|Y) p_{\bY|\bX}(Y|X) p_{\bX|\bu}(X|u) \nonumber \\
    & = \sum_{X \in \calT_u} p_{\bX|\bu}(X|u) \sum_{Y \in \calT_v} p_{\bY|\bX}(Y|X),
\end{align*}
and, therefore, may depend not only on $p_{\bY|\bX}$ (i.e., on $p_\bG$), but also on $p_{\bX|\bu}$. In the next theorem, we find the value of the rank transition probability for the case of a u.g.r.\ transfer matrix, and we show that it is independent of $p_{\bX|\bu}$. We also determine the channel transition probability in terms of the rank transition probability.

\medskip

\begin{thm}\label{thm:basic}
  The following holds for the MMC with u.g.r.\ transfer matrix:
  \begin{enumerate}
    \item Let $u$, $v$, and $r$ be nonnegative integers such that $r \leq \min \{ n, m \}$. We have
    \begin{equation} \label{eq:rank-transition}
      p_{\bv|\bu,\br}(v|u,r) = \frac {{u \brack v}_q} {{n \brack r}_q} {n-u \brack r-v}_q q^{v(n-u-r+v)},
    \end{equation}
    which does not depend on $p_{\bX|\bu}$. Thus, the rank transition probability is given by
    \[
      p_{\bv|\bu}(v|u) = \sum_r p_\br(r) p_{\bv|\bu,\br}(v|u,r),
    \]
    and the output rank probability is given by
    \[
      p_\bv(v) = \sum_u p_\bu(u) p_{\bv|\bu}(v|u).
    \]

    \item The channel transition probability is given by
    \begin{equation} \label{eq:transition-probability}
      p_{\bY|\bX}(Y|X)=
      \begin{cases} \displaystyle
        \frac {p_{\bv|\bu}(v|u)} {|\calT_v(\FF_q^{m \times u})|}, & \text{if }\langle Y \rangle \subseteq\langle X \rangle, \\
        0, & \text{else.}
      \end{cases}
    \end{equation}
    Moreover, if the input~$\bX$ is u.g.r., so is the output~$\bY$.
  \end{enumerate}
\end{thm}

\medskip

\begin{rem}
  Let $u$, $v$, and $r$ be nonnegative integers such that $r \leq \min \{ n, m \}$.  Recall from~\eqref{eq:gaussian-binomial} that the Gaussian binomial coefficient~${x \brack y}_q$ is nonzero if and only if $0 \leq y \leq x$. Thus, according to~\eqref{eq:rank-transition}, we have $p_{\bv|\bu,\br}(v|u,r) \neq 0$ if and only if $0 \leq v \leq u$ and $0 \leq r - v \leq n - u$; these, in turn, are equivalent to $u + r - n \leq v \leq \min \{ u, r \}$. This is expected: the upper bound follows trivially because $\rank GX \leq \min \{ \rank X, \rank G \}$, and the lower bound follows from Sylvester's rank inequality, which says that, if $G$ and $X$ are matrices of sizes $m \times n$ and $n \times \ell$, respectively, then $\rank X + \rank G - n \leq \rank GX$.
\end{rem}

\medskip

We next derive the channel capacity. We will see that u.g.r.\ input suffices to achieve the capacity, so that there is no need to consider more general inputs. Let
\begin{equation} \label{eq:mutual-information-ugr-def}
  I^*(p_\bu) \triangleq \max_{p_\bX : p_\bu} I(\bX ; \bY),
\end{equation}
where the maximum is over the collection of all matrix probability distributions $p_\bX$ with associated rank probability distribution equal to $p_\bu$, that is, over the set
\[
  \{ p_\bX: \textstyle{\sum_{X \in \calT_u}p_\bX(X) = p_\bu(u)}, \text{ for } u = 0, \ldots, n \}.
\]

\medskip

\begin{thm}\label{thm:capacity}
  The capacity of the MMC with u.g.r.\ transfer matrix is given by
  \[
    C = \max_{p_\bu} I^*(p_\bu),
  \]
  where $I^*(p_\bu)$, as defined in~\eqref{eq:mutual-information-ugr-def}, is achieved by u.g.r.\ input, and is given by
  \begin{equation} \label{eq:mutual-information-ugr-val}
    I^*(p_\bu) = \sum_v p_\bv(v) \log_q \frac {|\calT_v(\FF_q^{m \times \ell})|} {p_\bv(v)} - \sum_u h_u p_\bu(u),
  \end{equation}
  where
  \begin{equation} \label{eq:hu}
    h_u = \sum_v p_{\bv|\bu}(v|u) \log_q \frac{|\calT_v(\FF_q^{m \times u})|} {p_{\bv|\bu}(v|u)}.
  \end{equation}
\end{thm}

\medskip

From Theorem~\ref{thm:capacity}, we can see that the problem of finding the capacity and the corresponding optimal input for the MMC with u.g.r.\ transfer matrix, which was originally a convex optimization problem over $q^{n\ell}$ variables (namely, $p_\bX(X)$ for $X \in \FF_q^{n \times \ell}$), is simplified to another convex optimization problem, this time involving only $n+1$ variables (namely, $p_\bu(u)$, for $u = 0, \ldots, n$). The solution to this optimization problem can be obtained by standard methods (see, e.g.,~\cite{Boyd.04}).

We now focus on the special situation in which the input matrices are restricted to have \emph{constant rank}. This case is of interest for at least two reasons. First, constant-rank input happens to be asymptotically optimal both in the packet length and in the field size (as we shall see next). And second, most of the existing practical constructions for subspace codes are ``codes in the Grassmannian,'' that is, constant-dimension subspace codes~\cite{Koetter.Kschischang.08}.

Let $C_u$ denote the maximum channel mutual information when the input is restricted to rank-$u$ matrices. Let $u^*$ denote the value of $u$ that maximizes $C_u$, so that $C_{u^*} = \max_u C_u$. We call $C_u$ the \emph{rank-$u$ capacity}, and $C_{u^*}$ the \emph{constant-rank capacity} of the multiplicative finite-field matrix channel.

\medskip

\begin{thm}\label{thm:constant-rank}
  The rank-$u$ capacity of the MMC with u.g.r.\ transfer matrix is achieved by the uniform [over $\calT_u(\FF_q^{n \times \ell})$] input distribution, and is given by
  \begin{equation} \label{eq:capacity-constant-rank}
    C_u = \sum_v p_{\bv|\bu}(v|u) \log_q \frac {{\ell \brack v}_q} {{u \brack v}_q}.
  \end{equation}
   Moreover,
  \begin{equation} \label{eq:bounds}
    C_{u^*} \leq C \leq C_{u^*} + \log_q (\min \{ n, m \} + 1).
  \end{equation}
\end{thm}

\medskip

\begin{rem}
  In particular, if the input is always full rank (i.e., $\bu = n$), then $\bv = \br$ (since $\bv = \rank \bY = \rank \bG \bX = \rank \bG = \br$). The capacity becomes simply
  \[
    C_n = \sum_r p_\br(r) \log_q \frac {{\ell \brack r}_q} {{n \brack r}_q},
  \]
  a result obtained earlier in~\cite{Uchoa.10.arXiv}. Moreover, since $p_{\bv|\langle \bX \rangle}(v|U)$ only depends on~$U$ through $u = \dim U$ (see Theorem~\ref{thm:basic}), our result agrees with~\cite[Theorem~7]{Yang.10.arXiv}.
\end{rem}

\medskip

We next turn to the behavior of the channel for asymptotically large packet length~$\ell$, and asymptotically large field size~$q$. We show that, for both scenarios, constant-rank input suffices to achieve the capacity.

Consider first the asymptotic behavior in the packet length~$\ell$. In this situation, it is appropriate to define $\bar{C} \triangleq C / \ell$, the \emph{normalized capacity} of the matrix channel, measured in packets per channel use. We also define the \emph{normalized rank-$u$ capacity} as $\bar{C}_u \triangleq C_u / \ell$, and the \emph{normalized constant-rank capacity} as $\bar{C}_{u^*}$, where $u^*$ is the value of $u$ that maximizes~$\bar{C}_u$.

\medskip

\begin{thm} \label{thm:asymptotic-packet}
  Asymptotically in the packet length $\ell$, the normalized capacity of the MMC with u.g.r. transfer matrix is achieved with constant-rank uniform input, and is given by
  \[
    \lim_{\ell \to \infty} \bar{C} = E[\br].
  \]
  The optimal input rank is always $u^* = n$.
\end{thm}

\medskip

\begin{rem}
  This result is also obtained in~\cite[Corollary~1]{Yang.10.arXiv} for the case of an MMC with a general transfer matrix.
\end{rem}

\medskip

We now turn to the asymptotic behavior in the field size~$q$. In a general situation, the rank distribution may depend on $q$ [for example, the case in~\eqref{eq:rank-jafari}]. Thus, in what follows, we let
\[
  p^\infty_\br(r) \triangleq \lim_{q \to \infty} p_\br(r)
\]
denote the limiting distribution of $\br$ in $q$, assuming such a limit exists. Of course, when the rank distribution does not depend on $q$, then $p_\br^\infty(r) = p_\br(r)$.

\medskip

\begin{thm} \label{thm:asymptotic-field}
  Asymptotically in the field size $q$, the capacity of the MMC with u.g.r. transfer matrix is achieved with constant-rank uniform input, and is given by
  \[
    \lim_{q \to \infty} C = \max_u \left[ (\ell - u) \sum_r p_\br^\infty(r) \min \{ u, r \} \right].
  \]
\end{thm}

\medskip

\begin{rem}
   Consider random linear network coding in the absence of link errors and erasures. When the field size~$q$ is asymptotically large, it is known~\cite{Ho.06} that the transfer matrix will have rank $h$ with probability approaching one, where $h$ is the network mincut. In this case, $p_\br^\infty(r) = 1[r=h]$, so that
  \[
    \lim_{q \to \infty} C = \max_u \left[ (\ell - u) \min \{ u, h \} \right] = (\ell - u^*)u^*,
  \]
  where $u^* = \min \{h, \lfloor \ell/2 \rfloor \}$. For the sub-case in which $h = \min \{ n, m \}$, we have $u^* = \min \{ n, m, \lfloor \ell/2 \rfloor \}$, which agrees with~\cite[Proposition~3]{Silva.10} and \cite[Theorem~2]{Jafari.11}, since in both cases $p_\br^\infty(r) = 1[r = \min \{ n, m \}]$ [see equations~\eqref{eq:rank-silva} and~\eqref{eq:rank-jafari}].
\end{rem}

\medskip

Our last result is concerned with the optimality of subspace coding~\cite{Koetter.Kschischang.08} for the MMC with u.g.r.\ transfer matrix. Let $\calP(\FF_q^\ell, d)$ denote the set of all subspaces of~$\FF_q^\ell$ with dimension~$d$ or less.

\medskip

\begin{thm}\label{thm:subspace-channel}
  Consider the MMC with u.g.r.\ transfer matrix. Define $\bU \triangleq \langle \bX \rangle$ and $\bV \triangleq \langle \bY \rangle$. Then,
  \begin{equation} \label{eq:grouping-mutual-information}
    I(\bX;\bY) = I(\bU;\bV),
  \end{equation}
  for every input distribution $p_\bX$. Furthermore, for every $U \in \calP(\FF_q^\ell, n)$ and $V \in \calP(\FF_q^\ell, m)$, we have
  \begin{equation} \label{eq:grouping-transition-probability}
    p_{\bV|\bU}(V|U) = | \calT(\FF_q^{m \times \dim V}) | \ p_{\bY|\bX}(Y|X),
  \end{equation}
  where $X \in \FF_q^{n \times \ell}$ and $Y \in \FF_q^{m \times \ell}$ are any matrices such that $\langle X \rangle = U$ and $\langle Y \rangle = V$.
\end{thm}

\medskip

As a consequence of Theorem~\ref{thm:subspace-channel}, the matrix channel
\[
  (\calX = \FF_q^{n \times \ell}, \ p_{\bY|\bX}, \ \calY = \FF_q^{m \times \ell})
\]
can be transformed into a (simpler) \emph{subspace channel}
\[
  (\calU = \calP(\FF_q^\ell, n), \ p_{\bV|\bU}, \ \calV = \calP(\FF_q^\ell, m))
\]
with channel transition probability $p_{\bV|\bU}$ given by~\eqref{eq:grouping-transition-probability}. Concretely, the new channel is obtained by concatenating the original channel at the input with a device that takes a subspace $\bU$ to any matrix $\bX$ such that $\langle \bX \rangle = \bU$, and at the output with a device that computes $\bV = \langle \bY \rangle$. Due to~\eqref{eq:grouping-mutual-information}, any coding scheme for the matrix channel has a counterpart in the subspace channel achieving exactly the same mutual information, and vice versa. In particular, one may focus solely on $(\calU, p_{\bV|\bU}, \calV)$ when designing and analyzing capacity-achieving schemes.

\section{Proofs} \label{sec:proofs}

This section presents the proofs omitted from Section~\ref{sec:results}. In order to preserve space, we will often drop the subscripts of the probability distributions, writing, for example, $p(X)$ instead of $p_\bX(X)$. Before we proceed, we present a series of matrix enumeration results that will prove useful throughout this section.

\medskip

\begin{lem}\label{lem:subspace-superspace}
  Let $X \in \calT_u(\FF_q^{n \times \ell})$ be given. The number of matrices $Y \in \calT_v(\FF_q^{m\times \ell})$ such that $\langle Y \rangle \subseteq \langle X \rangle$ is given by
  \[
    |\{ Y \in \calT_v : \langle Y \rangle \subseteq \langle X \rangle \}|
      = |\calT_v(\FF_q^{m \times u})|.
  \]
  Now, let $Y \in \calT_v(\FF_q^{m \times \ell})$ be given. The number of matrices $X \in \calT_u(\FF_q^{n\times \ell})$ such that $\langle Y \rangle \subseteq \langle X \rangle$ is given by
  \[
    |\{ X \in \calT_u : \langle Y \rangle \subseteq \langle X \rangle \}|
      = |\calT_v(\FF_q^{m\times u})| \frac {|\calT_u(\FF_q^{n\times\ell})|} {|\calT_v(\FF_q^{m\times\ell})|}.
  \]
\end{lem}

\begin{IEEEproof}
  For every $X \in \calT_u(\FF_q^{n \times \ell})$, define
  \[
     \calJ(X) = \{ Y \in \calT_v : \langle Y \rangle \subseteq \langle X \rangle \}.
  \]
  Let $X_1, X_2 \in \calT_u(\FF_q^{n \times \ell})$. Then, there exist invertible matrices $S \in \FF_q^{n \times n}$ and $T \in \FF_q^{\ell \times \ell}$ such that $X_1 = SX_2T$. It is not hard to show that $Y \mapsto YT^{-1}$ is a bijection between $\calJ(X_1)$ and $\calJ(X_2)$, so that we must have $|\calJ(X_1)| = |\calJ(X_2)|$. Therefore, to compute the value of $|\calJ(X)|$, we can set
  \[
    X = \left[
    \begin{array}{cc}
      I_u & 0 \\
      0   & 0
    \end{array}
    \right] \in \FF_q^{n \times \ell},
  \]
  where $I_u$ is the $u \times u$ identity matrix. Since $Y \in \calJ(X)$ if and only if $Y$ is of the form $[ Y_0 ~~ 0 ]$, where $Y_0 \in \calT_v(\FF_q^{m \times u})$, we conclude that $|\calJ(X)| = |\calT_v(\FF_q^{m \times u})|$, as desired.

  Now, for every $Y \in \calT_v(\FF_q^{m \times \ell})$, define
  \[
     \calK(Y) = \{ X \in \calT_u : \langle Y \rangle \subseteq \langle X \rangle \}.
  \]
  Similarly to the previous paragraph, it is possible to show that $|\calK(Y_1)| = |\calK(Y_2)|$ for every $Y_1, Y_2 \in \calT_v(\FF_q^{m \times \ell})$. Consider then a bipartite graph where $X$s in $\calT_u(\FF_q^{n \times \ell})$ are the nodes in the left-hand side, $Y$s in $\calT_v(\FF_q^{m \times \ell})$ are the nodes in the right-hand side, and in which a node $X$ is connected with a node $Y$ if and only if $\langle Y \rangle \subseteq \langle X \rangle$. The number of edges connected with nodes in the left-hand side, namely,   $|\calT_u(\FF_q^{n \times \ell})| \, |\calJ(X)|$, must be equal to the number of edges connected with nodes in the right-hand side, namely, $|\calT_v(\FF_q^{m \times \ell})| \, |\calK(Y)|$, from which the second statement follows.
\end{IEEEproof}

\medskip

The next lemma is a combinatorial result by Brawley and Carlitz~\cite{Brawley.Carlitz.73}.

\medskip

\begin{lem}\label{lem:brawley-carlitz}
  Let $G_0 \in \calT_v(\FF_q^{m \times u})$ be a given matrix. The number of matrices $G \in \calT_r(\FF_q^{m \times n})$ whose left $m \times u$ sub-matrix is $G_0$ is given by
  \[
    \phi_q(m,n,u,r,v) \triangleq \frac {|\calT(\FF_q^{m \times r})|} {|\calT(\FF_q^{m \times v})|} {n - u \brack r - v}_q  q^{v(n - u - r + v)}.
  \]
\end{lem}

\medskip

We now derive another basic enumeration result which is closely related to the multiplicative finite-field matrix channel.

\medskip

\begin{lem}\label{lem:Y=GX}
  Let $X \in \calT_u(\FF_q^{n\times\ell})$ and $Y \in \calT_v(\FF_q^{m\times\ell})$. The number of matrices $G \in \calT_r(\FF_q^{m \times n})$ such that $GX = Y$ is
  \[
    |\{ G \in \calT_r : GX = Y \}| = \phi_q(m,n,u,r,v) \, 1[\langle Y \rangle \subseteq \langle X \rangle].
  \]
\end{lem}

\begin{IEEEproof}
  Let $X \in \calT_u(\FF_q^{n \times \ell})$, $Y \in \calT_v(\FF_q^{m \times \ell})$, and define
  \[
    \calJ(X,Y) = \{ G \in \calT_r : GX = Y \}.
  \]
  If $\langle Y\rangle \nsubseteq\langle X\rangle$, then clearly $|\calJ(X,Y)| = 0$, since no $G$ can take $X$ into $Y$. Suppose, then, that $\langle Y\rangle \subseteq\langle X\rangle$. Using a similar argument as employed in the proof of Lemma~\ref{lem:subspace-superspace}, we can conclude that it suffices to show the result for
  \[
    X = \left[
    \begin{array}{cc}
      I_u & 0\\
      0 & 0
    \end{array}
    \right] \in \FF_q^{n \times \ell},
  \]
  where $I_u$ is the $u \times u$ identity matrix. For this particular $X$, we must have $Y = [Y_0 ~~ 0]$ for some $Y_0 \in \calT_v(\FF_q^{m \times u})$ (recall that $\langle Y \rangle \subseteq \langle X \rangle$ is assumed). On the other hand, we also have $Y = GX = [G_0 ~~ 0]$, where $G_0 \in \FF_q^{m \times u}$ is the left $m \times u$ sub-matrix of $G$.  We thus have $G \in \calJ(X,Y)$ if and only if $G \in \calT_r(\FF_q^{m \times n})$ and $G_0 = Y_0 \in \calT_v(\FF_q^{m \times u})$. The result now follows from Lemma~\ref{lem:brawley-carlitz}.
\end{IEEEproof}

\medskip

We are finally ready to prove the theorems.

\medskip

\begin{IEEEproof}[Proof of Theorem~\ref{thm:basic}]
  Let $X \in \calT_u(\FF_q^{n \times \ell})$, $Y \in \calT_v(\FF_q^{m \times \ell})$, and $r$ such that $0 \leq r \leq \min \{ n, m \}$. We have
  \begin{align*}
    p(Y|X,r)
      & = \sum_{G \in \calT_r} p(G|r) \, p(Y|X,G) \\
      & \overset{(a)} = \frac 1 {|\calT_r(\FF_q^{m \times n})|} \sum_{G \in \calT_r} 1 [Y = GX] \\
      & \overset{(b)} = \frac 1 {|\calT_r(\FF_q^{m \times n})|} \phi_q(m,n,u,r,v) \, 1[\langle Y \rangle \subseteq \langle X \rangle],
  \end{align*}
  where $(a)$ follows because $\bG$ is u.g.r., and $(b)$ follows from Lemma~\ref{lem:Y=GX}.  Therefore, from Lemma~\ref{lem:subspace-superspace}, we may write
  \[
    p(v|X,r) = \sum_{Y \in \calT_v} p(Y|X,r) = \frac {|\calT_v(\FF_q^{m \times u})|} {|\calT_r(\FF_q^{m \times n})|} \phi_q(m,n,u,r,v),
  \]
  so that,
  \begin{align*}
    p(v|u,r)
      & = \sum_{X \in \calT_u} p(X|u) \, p(v|X,r) \\
      & = \sum_{X \in \calT_u} p(X|u) \frac {|\calT_v(\FF_q^{m \times u})|} {|\calT_r(\FF_q^{m \times n})|} \phi_q(m,n,u,r,v) \\
      & = \frac {|\calT_v(\FF_q^{m \times u})|} {|\calT_r(\FF_q^{m \times n})|} \phi_q(m,n,u,r,v),
  \end{align*}
  and~\eqref{eq:transition-probability} follows by comparing the expressions for $p(Y|X,r)$ and $p(v|u,r)$. To prove~\eqref{eq:rank-transition}, we substitute $\phi_q(m,n,u,r,v)$ with its definition (see Lemma~\ref{lem:brawley-carlitz}), to get
  \begin{align*}
    p(v|u,r)
      & = \frac {|\calT_v(\FF_q^{m \times u})|} {|\calT_r(\FF_q^{m \times n})|} \frac {|\calT(\FF_q^{m \times r})|} {|\calT(\FF_q^{m \times v})|} {n - u \brack r - v}_q  q^{v(n - u - r + v)} \\
      & = \frac {{u \brack v}_q} {{n \brack r}_q} {n-u \brack r-v}_q q^{v(n-u-r+v)},
  \end{align*}
  where we used~\eqref{eq:chi-rank} in the last step.

  To finish the proof, assume that $\bX$ is u.g.r.  Then, for each $Y \in \calT_v(\FF_q^{m \times \ell})$, we have
  \begin{align*}
    p(Y)
      & = \sum_u \sum_{X \in \calT_u} p(Y|X) p(X) \\
      & \overset{(a)} = \sum_u \frac {p(u)} {|\calT_u(\FF_q^{n \times \ell})|} \sum_{X \in \calT_u} p(Y|X) \\
      & \overset{(b)} = \sum_u \frac {p(u)} {|\calT_u(\FF_q^{n \times \ell})|} \frac {p(v|u)} {|\calT_v(\FF_q^{m \times u})|} \sum_{X \in \calT_u} 1 [\langle Y \rangle \subseteq\langle X \rangle] \\
      & \overset{(c)} = \sum_u \frac {p(u)} {|\calT_u(\FF_q^{n \times \ell})|} \frac {p(v|u)} {|\calT_v(\FF_q^{m \times u})|} |\calT_v(\FF_q^{m\times u})| \frac {|\calT_u(\FF_q^{n\times\ell})|} {|\calT_v(\FF_q^{m\times\ell})|} \\
      & = \frac {p(v)} {|\calT_v(\FF_q^{m\times\ell})|},
  \end{align*}
  where $(a)$ follows because $\bX$ is u.g.r., $(b)$ follows from \eqref{eq:transition-probability}, and $(c)$ follows from Lemma~\ref{lem:subspace-superspace}. Therefore, $\bY$ is also u.g.r., as claimed.
\end{IEEEproof}

\medskip

\begin{IEEEproof}[Proof of Theorem~\ref{thm:capacity}]
  For each $X \in \calT_u(\FF_q^{n \times \ell})$, we have
  \begin{align*}
    H(\bY|\bX = X)
      & = \sum_v \sum_{Y \in \calT_v} p(Y|X) \log_q \frac 1 {p(Y|X)} \\
      & = \sum_v p(v|u) \log_q \frac{|\calT_v(\FF_q^{m \times u})|} {p(v|u)} = h_u,
  \end{align*}
  where we substituted $p(Y|X)$ as in~\eqref{eq:transition-probability}. Averaging over all $X \in \FF_q^{n \times \ell}$, we get
  \begin{align*}
    H(\bY|\bX)
      & = \sum_u \sum_{X \in \calT_u} H(\bY|\bX = X) p(X) \\
      & = \sum_u h_u \sum_{X \in \calT_u} p(X) \\
      & = \sum_u h_u p(u),
  \end{align*}
  which depends on $p_\bX$ only through $p_\bu$. Therefore,
  \begin{align*}
    I^*(p_\bu)
      & = \max_{p_\bX : p_\bu} I(\bX ; \bY) \\
      & = \max_{p_\bX : p_\bu} [ H(\bY) - H(\bY | \bX) ] \\
      & = [ \max_{p_\bX : p_\bu} H(\bY) ] - \sum_u h_u p(u),
  \end{align*}
  and we get the desired result from~\eqref{eq:entropy-bound}.
\end{IEEEproof}

\medskip

\begin{IEEEproof}[Proof of Theorem~\ref{thm:constant-rank}]
  If the input is restricted to rank-$u$ matrices, then $\bu = u$ is a constant, and therefore $p(v) = p(v|u)$. The channel mutual information given by Theorem~\ref{thm:capacity} simplifies to
  \[
    \sum_v p(v|u) \log_q \frac {|\calT_v(\FF_q^{m \times \ell})|} {|\calT_v(\FF_q^{m \times u})|},
  \]
  and we get~\eqref{eq:capacity-constant-rank} by applying~\eqref{eq:chi-rank}.

  The lower bound of~\eqref{eq:bounds} is immediate. Similarly to Yang~{\it et al.} in~\cite[Lemma~4]{Yang.10.arXiv}, we can rewrite the mutual information~\eqref{eq:mutual-information-ugr-val} as
  \begin{align*}
    I^*(p_\bu)
     & = \sum_v p(v) \log_q \frac {|\calT_v(\FF_q^{m \times \ell})|} {p(v)} - \sum_u p(u) h_u \\
     & = \sum_{u,v} p(u) p(v|u) \log_q \frac {|\calT_v(\FF_q^{m \times \ell})|} {p(v)} \ +  \\
       & \quad - \sum_{u,v} p(u) p(v|u) \log_q \frac{|\calT_v(\FF_q^{m \times u})|} {p(v|u)} \\
     & = \sum_{u,v} p(u) p(v|u) \log_q \frac {|\calT_v(\FF_q^{m \times \ell})|} {|\calT_v(\FF_q^{m \times u})|} \ + \\
       & \quad + \sum_{u,v} p(u) p(v|u) \log_q \frac {p(v|u)} {p(v)} \\
       & = \sum_u p(u) C_u + I(\bu;\bv),
  \end{align*}
  where $I(\bu;\bv)$ is the mutual information between the random variables $\bu$ and $\bv$. The upper bound of~\eqref{eq:bounds} then follows because $\sum_u p(u) C_u \leq \max_u C_u = C_{u^*}$ and $I(\bu;\bv) \leq \log_q (\min \{ n, m \} + 1)$.
\end{IEEEproof}

\medskip

\begin{IEEEproof}[Proof of Theorem~\ref{thm:asymptotic-packet}]
  Dividing~\eqref{eq:bounds} by $\ell$, and taking the limit when $\ell \to \infty$, we obtain
  \[
    \lim_{\ell \to \infty} \bar{C} = \lim_{\ell \to \infty}\bar{C}_{u^*},
  \]
  so that constant-rank input is sufficient to achieve capacity for asymptotically large $\ell$. Now, dividing~\eqref{eq:capacity-constant-rank} by $\ell$, and taking the limit when $\ell \to \infty$, we obtain
  \begin{align*}
    \lefteqn{\lim_{\ell \to \infty} \bar{C}_u = \sum_v p(v|u) \left( \lim_{\ell \to \infty} \frac 1 \ell \log_q \frac {{\ell \brack v}_q} {{u \brack v}_q} \right)} \\
      & \quad = \sum_v p(v|u) \left( \lim_{\ell \to \infty} \frac 1 \ell \log_q {\ell \brack v}_q - \lim_{\ell \to \infty} \frac 1 \ell \log_q {u \brack v}_q \right) \\
      & \quad = \sum_v p(v|u) \left( \lim_{\ell \to \infty} \frac 1 \ell \log_q {\ell \brack v}_q \right) \\
      & \quad = \sum_v v \, p(v|u) = E[\bv | \bu = u],
  \end{align*}
  where the first equality in the last line is a consequence of~\eqref{eq:gaussian-binomial-bounds}. Finally, since $\bv \leq \br$, we have
  \[
    E[\bv | \bu = u, \br = r] \leq r = E[\bv | \bu = n, \br = r],
  \]
  for all $u \in \{ 0, \ldots, n \}$. Multiplying both sides by $p(r)$ and summing over $r$, we obtain
  \[
    E[\bv | \bu = u] \leq E[\br] = E[\bv | \bu = n],
  \]
  which shows that $\lim_{\ell \to \infty} \bar{C}_u = E[\bv | \bu = u]$ is maximum when $u = n$, with the maximum value being $E[\br]$.
\end{IEEEproof}

\medskip

For the next result, we will need the following intuitive fact.

\medskip

\begin{lem} \label{lem:rank-transition-asympt}
  We have
  \[
    \lim_{q \to \infty} p(v|u,r) =
    \begin{cases}
      1, & \text{if } v = \min \{ u, r \}, \\
      0, & \text{else.}
    \end{cases}
  \]
\end{lem}

\begin{IEEEproof}
  This is clearly true if $v > \min \{ u, r \}$. When $v \leq \min\{u,r\}$, we have from~\eqref{eq:gaussian-binomial-bounds} and from Theorem~\ref{thm:basic} that
  \begin{multline*}
    q^{v(u - v)} \cdot \gamma_q^{-1} q^{-r(n - r)} \cdot q^{(r-v)(n-u-r+v)} \cdot q^{v(n-u-r+v)} \\
      \leq p(v|u,r) = {u \brack v}_q {n \brack r}_q^{-1} {n-u \brack r-v}_q q^{v(n-u-r+v)} \leq \\
      \gamma_q q^{v(u - v)} \cdot q^{-r(n - r)} \cdot \gamma_q q^{(r-v)(n-u-r+v)} \cdot q^{v(n-u-r+v)}.
  \end{multline*}
  After simplifying, we get
  \begin{equation*}
    \gamma_q^{-1} q^{-(u - v)(r - v)} \leq p(v|u,r) \leq \gamma_q^2 q^{-(u - v)(r - v)},
  \end{equation*}
  and the desired result follows because $\lim_{q \to \infty} \gamma_q = 1$.
\end{IEEEproof}

\medskip

\begin{IEEEproof}[Proof of Theorem~\ref{thm:asymptotic-field}]
  The quantity $\log_q(\min \{ n, m\} + 1)$ in the right-hand side of~\eqref{eq:bounds} goes to zero as~$q \to \infty$, so that
  \[
    \lim_{q \to \infty} C = \lim_{q \to \infty} C_{u^*},
  \]
  that is, constant-rank input suffices for asymptotically large $q$. Now, from~\eqref{eq:capacity-constant-rank}, we have
  \[
    \lim_{q \to \infty} C_u = \sum_v \left( \lim_{q \to \infty} p(v|u) \right) \left( \lim_{q \to \infty} \log_q \frac {{\ell \brack v}_q} {{u \brack v}_q} \right)
  \]
  For the first parenthesis, we have from Lemma~\ref{lem:rank-transition-asympt} that
  \[
    \lim_{q \to \infty} p(v|u) = \sum_{r=0}^n p_\br^\infty(r) 1[ v=\min \{u,r\}].
  \]
  For the second parenthesis, we have from~\eqref{eq:gaussian-binomial-bounds} that
  \[
    \lim_{q \to \infty} \log_q\frac{{\ell \brack v}_q}{{u \brack v}_q} = v(\ell - u).
  \]
  Therefore,
  \begin{align*}
    \lim_{q \to \infty} C_u
      & = \sum_v \sum_r p_\br^\infty(r) 1[ v = \min \{ u ,r \} ]  v (\ell - u) \\
      & = (\ell - u) \sum_r p_\br^\infty(r) \sum_v 1[ v=\min \{ u,r \} ] \ v \\
      & = (\ell - u) \sum_r p_\br^\infty(r) \min \{ u,r \},
  \end{align*}
  as desired.
\end{IEEEproof}

\medskip

\begin{IEEEproof} [Proof of Theorem~\ref{thm:subspace-channel}]
  From Theorem~\ref{thm:basic} we know that $p_{\bY|\bX}(Y|X)$ depends on $X$ and $Y$ only through $\langle X \rangle$ and $\langle Y \rangle$. Therefore, according to Lemma~\ref{lem:grouping}, the maps $f(\bX) = \langle \bX \rangle$ and $g(\bY) = \langle \bY \rangle$ are information-lossless. This proves~\eqref{eq:grouping-mutual-information}.

  To prove~\eqref{eq:grouping-transition-probability}, we first apply the input grouping to the original matrix channel  $(\calX, p_{\bY|\bX}, \calY)$, to get an intermediate channel $(\calU, p_{\bY|\bU}, \calY)$, with $p_{\bY|\bU}(Y|U) = p_{\bY|\bX}(Y|X)$, where $X$ is such that $\langle X \rangle = U$.  Then, we apply the output grouping to this intermediate channel to get the subspace channel $(\calU, p_{\bV|\bU}, \calV)$ with
  \begin{align*}
    p_{\bV|\bU}(V|U)
      & = \sum_{Y' : \langle Y' \rangle = V} \ p_{\bY|\bU}(Y'|U) \\
      & = | \calT(\FF_q^{m \times \dim V}) | \ p_{\bY|\bU}(Y|U),
  \end{align*}
  where $Y$ is such that $\langle Y \rangle = V$. Note that the last step in the above equation follows from
  \[
    |\{ Y' \in \FF_q^{m \times \ell} : \langle Y' \rangle = V \} | = | \calT(\FF_q^{m \times \dim V}) |,
  \]
  which is true because associated with every $Y' \in \FF_q^{m \times \ell}$ such that $\langle Y' \rangle = V $, there is a unique full-rank matrix $T \in \calT(\FF_q^{m \times \dim V})$ such that $Y' = T \tilde{Y}$, where $\tilde{Y} \in \calT(\FF_q^{\dim V \times \ell})$ is any fixed full-rank matrix satisfying $\langle \tilde{Y} \rangle = V$.
\end{IEEEproof}

\section{Conclusions}\label{sec:conclusion}

This work has considered probabilistic multiplicative finite-field matrix channels in which the transfer matrix is uniformly distributed conditioned on its rank. We advocate the application of this channel model in practical noncoherent network coding systems subject to link erasures, for we believe it is flexible enough to capture the essential characteristics of the system, while still being mathematically tractable. This contrasts with previously considered channel models, which are either too restrictive or too complex.

As contributions, we have shown that the problem of finding the channel capacity can be reduced to a convex optimization problem on $n+1$ variables (rather than $q^{n \ell}$), allowing for easy numerical computation by standard techniques. We have also specialized our results to the important case of constant-rank input, in which we were able to find a closed-form expression for the capacity. For asymptotically large field or packet length, we have shown that constant-rank input is optimal. Finally, we have proven that even in our more general setup, subspace coding is still sufficient to achieve capacity. Many of our results generalize existing conclusions in prior literature.

The present paper has focused mainly on the capacity and mutual information of the multiplicative finite-field matrix channel. The design of low-complexity capacity-achieving schemes for this channel is an important and still largely open problem. Recent work by Yang~{\it et al.}~\cite{Yang.10.arXiv,Yang.10.ISIT} has addressed this problem by considering the construction of codes based on the expected value of the rank of the transfer matrix,~$E[\br]$. Nevertheless, the design of codes based on the rank distribution~$p_\br$ is yet to be investigated.  Finally, another challenging and interesting research line motivated by the present work is the computation of the rank distribution as a function of a given network topology.

\appendices
\section{A Variation of the Crypto Lemma\protect\footnote{This appendix is a joint work with Chen Feng.}}\label{ape:crypto}

We start by recalling the following well-known result, known as the \emph{crypto lemma} for the case of finite groups~\cite{Forney.03.Allerton}.

\medskip

\begin{lem} \label{lem:crypto-0}
  Let $(\calG, \cdot)$ be a finite group. Let $\by = \bg \cdot \bx$, where $\bx$ and $\bg$ are random variables over~$\calG$, and $\bg$ is uniform over~$\calG$ and independent of~$\bx$.  Then, $\by$ is uniform over~$\calG$ and independent of~$\bx$.
\end{lem}

\medskip

Now, let $\calS$ be a set. Recall that a \emph{(left) group action} of~$\calG$ on~$\calS$ is a binary operator $\circ : \calG \times \calS \to \calS$ such that $(g_1 \cdot g_2) \circ x = g_1 \circ (g_2 \circ x)$, for all $g_1, g_1 \in \calG$ and $x \in \calS$; and $e \circ x = x$, for all $x \in \calS$, where $e$ is the identity element of $\calG$. Every group $\calG$ acts on itself ($\calS = \calG$) by left multiplication, that is, through the action given by $g \circ x = g \cdot x$. This appendix generalizes the crypto lemma from this special case to the case of an arbitrary action of $\calG$ on some finite set~$\calS$.  Before we proceed, we need to recall a few basic facts about group actions~\cite[\S 4.1]{Dummit.04}.

For every $x \in \calS$, the \emph{orbit} of $\calG$ containing $x$ is defined as $\calG \circ x \triangleq \{ g \circ x : g \in \calG \}$. The relation on $\calS$ defined by
\begin{equation*}
  \text{$x \sim y$} \qquad \text{iff} \qquad \text{$x = g \circ y$ for some $g \in \calG$}
\end{equation*}
is an equivalence relation. We have $x \sim y$ iff $\calG \circ x = \calG \circ y$ iff $x$ and $y$ are in the same orbit. The size of each orbit is given by $| \calG \circ x | = |\calG| / |\calG_{x,x}|$, where $\calG_{x,x} \triangleq \{ g \in \calG : g \circ x = x \}$ is the \emph{stabilizer} of $x$ in $\calG$ (a subgroup of $\calG$). An action is called \emph{transitive} if there is only one orbit.

\medskip

\begin{lem}\label{lem:crypto-1}
  Let $(\calG, \cdot)$ be a finite group, $\calS$ a finite set, and $\circ : \calG \times \calS \to \calS$ a group action of~$\calG$ on~$\calS$. Let $\by = \bg \circ \bx$ (so that $\bx$ and $\by$ lie in the same orbit), where $\bx$ and $\bg$ are random variables over~$\calS$ and $\calG$, respectively, and~$\bg$ is uniform over~$\calG$ and independent of~$\bx$.  Then, $\by$ is piece-wise uniform over the orbits of the action and conditionally independent of $\bx$ given that a particular orbit occurs.
\end{lem}

\begin{rem}
  In particular, if the action is transitive, then $\by$ is uniform over $\calS$ and independent of $\bx$.  This is the case of the action $g \circ x = g \cdot x$, so we recover Lemma~\ref{lem:crypto-0}.
\end{rem}

\begin{IEEEproof}
  Since $\bg$ is uniform and independent of $\bx$, we have that, for all $x,y \in \calS$,
  \begin{equation*}
    p_{\by|\bx}(y|x) = \frac{|\calG_{x,y}|}{|\calG|},
  \end{equation*}
  where $\calG_{x,y} \triangleq \{ g \in \calG : g \circ x = y \}$. If $x \sim y$ (so that $\calG \circ x = \calG \circ y$), it can be shown that $\calG_{x,y}$ is a coset of the stabilizer~$\calG_{x,x}$, which implies $|\calG_{x,y}| = |\calG_{x,x}|$, and thus
  \begin{equation*}
    p_{\by|\bx}(y|x) = \frac{|\calG_{x,x}|}{|\calG|} = \frac{1}{|\calG \circ x|} = \frac{1}{|\calG \circ y|}.
  \end{equation*}
  On the other hand, if $x \nsim y$, then clearly $p_{\by|\bx}(y|x) = 0$. Therefore,
  \begin{align*}
    p_\by(y)
      & = \sum_x p_{\by|\bx}(y|x) p_\bx(x) \\
      & = \frac{1}{|\calG \circ y|} \sum_{x : x \sim y} p_\bx(x) \\
      & = \frac{\Pr[\bx \sim y]}{|\calG \circ y|} = \frac{\Pr[\by \sim y]}{|\calG \circ y|},
  \end{align*}
  and the lemma follows.
\end{IEEEproof}

\medskip

Theorem~\ref{thm:worst-case} is a corollary of this result.

\medskip

\begin{IEEEproof}[Proof of Theorem~\ref{thm:worst-case}]
  The result follows after applying Lemma~\ref{lem:crypto-1} with $\calG = \calT(\FF_q^{m \times m}) \times \calT(\FF_q^{n \times n})$, where the operation is $(T'_1, T'_2) \cdot (T_1, T_2) = (T'_1 T_1, T_2 T'_2)$, $\calS = \FF_q^{m \times n}$, and $\circ : \calG \times \calS \to \calS$ defined by $(T_1, T_2) \circ M = T_1 M T_2$. The facts that $(\calG, \cdot)$ is a group and $\circ$ is an action of $\calG$ on $\calS$ follow from basic linear algebra;  the orbits, in this case, are $\{ \calT_r(\FF_q^{m \times n}) : r=0,\ldots,\min\{n,m\} \}$, which are completely characterized by the rank of $\bG$.
\end{IEEEproof}

\section*{Acknowledgments}

The authors would like to thank Chen Feng, Frank Kschischang, and Shenghao Yang for useful discussions. We are also thankful for the anonymous reviewers for their helpful comments and suggestions.

\bibliographystyle{IEEEtran}
\bibliography{biblio}

\end{document}